# Observation of topological polaritons and photonic *magic angles* in twisted van der Waals bi-layers

Guangwei Hu[1, 2, †], Qingdong Ou[3, †], Guangyuan Si[4], Yingjie Wu[3], Jing Wu[5], Zhigao Dai[3, 6], Alex Krasnok[2], Yarden Mazor[7], Qing Zhang[1], Qiaoliang Bao[3, *], Cheng-Wei Qiu[1, *], Andrea Alù[2,7,8, *]

[1]Department of Electrical and Computer Engineering, National University of Singapore, Singapore, Singapore.
[2]Photonics Initiative, Advanced Science Research Center, City University of New York, New York, NY, USA.
[3]Department of Materials Science and Engineering, and ARC Centre of Excellence in Future Low-Energy Electronics Technologies (FLEET), Monash University, Clayton, Victoria, Australia.
[4]Melbourne Centre for Nanofabrication, Victorian Node of the Australian National Fabrication Facility, Clayton, VIC, Australia.
[5]Institute of Materials Research and Engineering, A*STAR (Agency for Science, Technology and Research), Singapore, Singapore.
[6]Faculty of Materials Science and Chemistry, China University of Geosciences, 388 Lumo Road, Wuhan 430074, P. R. China.
[7]Department of Electrical and Computer Engineering, The University of Texas at Austin, Austin, TX, USA.
[8]Physics Program, Graduate Center, City University of New York, New York, NY, USA.

* Correspondence to: aalu@gc.cuny.edu; chengwei.qiu@nus.edu.sg; qiaoliang.bao@gmail.com
† these authors contributed equally: Guangwei Hu, Qingdong Ou.

**Twisted two-dimensional bi-layers offer exquisite control on the electronic bandstructure through the interlayer rotation and coupling, enabling magic-angle flat-band superconductivity and moiré excitons. Here, we demonstrate how analogous principles, combined with large anisotropy, enable extreme control and manipulation of the photonic dispersion of phonon polaritons (PhPs) in van der Waals (vdW) bi-layers. We experimentally observe tunable topological transitions from open (hyperbolic) to closed (elliptic) dispersion contours in twisted bi-layered α-MoO₃ at photonic magic angles, induced by polariton hybridization and robustly controlled by a topological quantity. At these transitions the bilayer dispersion flattens, exhibiting low-loss tunable polariton canalization and diffractionless propagation with resolution below $\lambda_0/40$. Our findings extend twistronics and moiré physics to nanophotonics and polaritonics, with great potential for nano-imaging, nanoscale light propagation, energy transfer and quantum applications.**

Twisted stacks of two-dimensional (2D) materials forming vertical heterostructures have raised tremendous interest in recent years, as they have been shown to support superconductivity at magic rotation angles that induce a flat Fermi surface in bilayer graphene[1], topological excitons due to twisted-angle-dependent interlayer hopping in bilayer transition-metal dichalcogenides[2-7], and interlayer magnetism[8], among several exotic electronic phenomena. These unusual electronic responses emerge from hybridization and the emergence of moiré superlattices controlled by the rotation angle, with powerful opportunities in the developing field of *twistronics*[9]. The extension of these concepts to photonics has recently been explored in atomically-thin photonic crystals in bilayer graphene[10,11] and in twisted hexagonal boron nitride[12].

We have recently theoretically proposed that a twisted stack of hyperbolic metasurfaces, each formed by densely packed graphene nanoribbons supporting hyperbolic plasmons[13], may enable an unusual control of the plasmon dispersion, analogous to moiré physics but for photons[14]. However, the extreme anisotropy of these artificial metastructures, at the basis of their hyperbolic response, is fundamentally limited by their granularity, inducing strong nonlocality[13-16] that hinders the practical verification of these concepts. In-plane hyperbolicity arises in metasurfaces when the imaginary part of the effective surface impedance along two orthogonal transverse directions have different signs[17-18], leading to sub-diffractive collimated surface wave propagation and enhanced local density of states[13]. Phonon polaritons (PhPs), quasiparticles stemming from collective oscillations between photons and lattice vibrations, can be naturally endowed with a hyperbolic response in polar van der Waals (vdW) nanomaterials[19-24]. While most of these naturally occurring hyperbolic polaritons propagate out of plane, it has recently been shown that in-plane hyperbolic PhPs arise in α-phase molybdenum trioxide (α-MoO$_3$) flake[24-26]. Within their Reststrahlen band (RB) from 818 cm$^{-1}$ to 974 cm$^{-1}$, the real part of permittivity of α-MoO$_3$ is negative along the [100] direction but positive along the [001] direction. Such extreme anisotropy allows in-plane low-loss hyperbolic PhP propagation, as experimentally revealed in recent works[24, 26].

In this work, we experimentally demonstrate that these unusual optical features enable extreme control of the polariton dispersion through the interlayer coupling in twisted bi-layered (tBL) α-MoO$_3$ flakes, due to the emergence of topological PhPs. The topological nature of this phenomenon emerges in two ways: first, their dispersion contours change dramatically, from hyperbolic (open) to elliptic (closed) dispersion lines, as a function of the rotation angle, yielding a photonic topological transition[27] governed by the interlayer coupling of the in-plane hyperbolic PhPs supported by each layer individually. In addition, this transition is determined by a topological quantity, i.e., the number of anti-crossing points (denoted as $N_{ACP}$) of the dispersion lines of each isolated layer in reciprocal space, with the isofrequency contours playing the role of a Fermi surface for electrons, analogous to a *Lifshitz transition* in electronics[27, 28]. The inherently robust nature of these transitions enables our experimental observation in several samples using real-space nanoimaging[29]. At the rotation angle where the topological transition arises (termed as *topological transition magic angle* hereafter), the dispersion necessarily flattens to change its topology from closed to open, yielding diffractionless and low-loss directional PhP canalization[30, 31]. Our studies provide important perspectives on topological polaritons, enhanced light-matter interactions and extreme photonic dispersion engineering with the tools of twistronics, for opportunities spanning from bio-sensing, imaging, near-field energy transfer to quantum nanophotonics.

The tBL α-MoO$_3$ geometry under consideration is shown in Fig. 1a. The top and bottom layers have thickness $d_1$ and $d_2$, respectively. The global Cartesian coordinate system is defined such that the *x* and *y* axes are along the [100] and [001] crystal directions of the bottom α-MoO$_3$ flake; Δθ (∈[-90˚, 90˚]) is the rotation angle of the [100] crystal direction of the top flake with respect to the *x*-axis, positive in the anti-clockwise direction. A representative image of a tBL α-MoO$_3$ sample is shown in Fig. 1b. To model and analytically study the system, we examine the changes in dispersion introduced by rotation solving source-free Maxwell equations, treating each α-MoO$_3$ flake as a 2D conductivity sheet. In this model, we can reasonably assume that the two surfaces are separated by an infinitesimally thin dielectric spacer to take evanescent hybridization into account (see method). For a single layer, hyperbolic PhPs propagate at the open angle ($\beta = \mathrm{atan}\left(\sqrt{-\varepsilon_{11}/\varepsilon_{22}}\right)$, where $\varepsilon_{11}$ ($\varepsilon_{22}$) is the permittivity component along the [100] ([001]) crystal direction (Fig. S1). For tBL flakes, the two hyperbolic bands of the individual layers strongly couple to each other at the points in reciprocal space where they cross, leading to anti-crossing in their dispersion. Interestingly, the number of anti-crossing points, $N_{ACP}$, in reciprocal space directly determines the hyperbolic or elliptic nature of the isofrequency contour of tBL α-MoO$_3$[14]. In particular, when $N_{ACP}$=2 the bilayer dispersion will remain hyperbolic, whereas for $N_{ACP}$=4 it turns elliptical. The integer number of anti-crossing is simply determined, using topological arguments, by the rotation angle (Δθ), which rotates one of the bands with respect to the other, and the open angle (β) of the hyperbolic dispersion band of each isolated layer. As a result, $N_{ACP}$ provides control over the bilayer dispersion with inherent topological robustness (see supplementary information). As summarized in Fig. 1c, in the frequency range from 860 cm$^{-1}$ to 940 cm$^{-1}$, $β > 45°$ (Fig. S1) and we expect hyperbolic dispersion ($N_{ACP}$=2) when $|Δθ| < |180° − 2β|$, and elliptic dispersion ($N_{ACP}$=4) when $|Δθ| > |180° − 2β|$ (Fig. S2). Fig. 1d specifically illustrates an example of different dispersion regimes with respect to $Δθ$ at fixed frequency ω=925.9 cm$^{-1}$, as numerically verified by full-wave simulations in Fig. 1e-i. A point dipole spaced 200 nm over the tBL α-MoO$_3$ flake ($d_1$=$d_2$=100 nm) is used to excite PhPs, with fields monitored 20 nm over the uppermost surface (see methods and Fig. S3). The electric field pattern (top panels, Fig. 1e-i) turns from hyperbolic to elliptical when the rotation angle increases, reflected in the calculated Fourier spectrum (magnitude plot in bottom panels, Fig. 1e-i) and in the analytical dispersion (red solid lines in bottom panels, Fig. 1e-i), with a topological transition from open to closed at the critical transition angle ($|180° − 2β|$). At this rotation angle, the dispersion flattens and the field becomes highly directional (Fig. 1g, h), which is reminiscent of the flat Fermi surface responsible for magic-angle superconductivity in bilayer graphene[1]. Thus, we term this angle as a *topological transition magic angle*. Importantly, this extreme form of dispersion engineering and the closed or open nature of the resulting bands does not require that the layers are identical to each other, and it is strongly robust to disorder, as long as the integer $N_{ACP}$ does not change, revealing inherent robustness of these topological transitions.

In order to validate our predictions, we used mechanical exfoliation to obtain individual α-MoO$_3$ flakes, and then transferred one flake onto the other with deterministic alignment and stacking order (see method). Three samples are shown in Fig. 2b-d, with rotation angles Δθ=-44˚, 65˚, and -77˚. In order to experimentally observe the rotation-induced polaritonic dispersion engineering, we used infrared (IR) real-space nanoimaging based on scattering-type scanning near-field optical microscopy (*s*-SNOM), as shown in Fig. 2a. Since PhPs have a significant momentum mismatch compared to free-space radiation, it is challenging to excite and image them in far field. A resonant Au antenna or Ag nanowire has been previously used to excite PhPs[26], and a metallic atomic force microscopy (AFM) tip assembled in s-SNOM has been used to couple out and map PhPs. However, using chemical solution based lithography to fabricate metallic nanostructures may introduce damage and contamination on α-MoO$_3$.[32] For this reason, we used focused ion beam to make a point defect (PD) in the form of a hole in the sample (black dots, Fig. 2b-d) and excite the PhPs using a metallic AFM tip. The tip-launched PhPs reflect at the PD and produce polariton interference, yielding the near-field images (Fig. S4).

As a benchmark, the PhPs supported by a single α-MoO$_3$ layer (top layer of the tBL α-MoO$_3$ sample of Δθ=-77˚) were first probed, clearly showing a hyperbolic wavefront (Fig. 2e), further confirmed by its hyperbolic Fourier spectrum (Fig. 2i), validating our PD-assisted excitation/imaging technique. Next we analyzed at the same frequency the response of the three tBL. The measured wavefront and retrieved dispersion curves preserve the hyperbolic features in the tBL α-MoO$_3$ sample for Δθ=-44˚ (Fig. 2f, j). Such hyperbolicity reduces to nearly straight PhPs wavefront for Δθ=65˚ (Fig. 2g, k), yielding a topological transition, and finally turns into elliptical wavefronts for Δθ=-77˚ (Fig. 2h, l). The observed degree of tunability of the dispersion in these twisted bilayers is remarkable, considering that individually each of them supports just hyperbolic polaritons in this frequency range. These observations provide clear evidence of our theoretically predicted twist-induced topological transition and extreme dispersion engineering controlled by $N_{ACP}$. Our experimental results are also numerically supported by our simulations with excellent agreement (Fig. S5). The experimentally measured dispersion fits analytical and numerical curves with a factor of two discrepancy in wavenumber, as expected due to the fact that we measured standing wave PhPs reflected by PD (Fig. S4 and Fig. S5). All measurements in Fig. 2 correspond to 903.8 cm$^{-1}$, but similar topological transitions have been observed at other frequencies/angles, consistent with Fig. 1c, and directly controlled by $N_{ACP}$ in reciprocal space (see results at 925.9 cm$^{-1}$ in Fig. S6).

The topological transition magic angles are controlled by the hyperbolic dispersion of each layer, as summarized in Fig. 1c. Four samples with different rotation angles (Δθ=-44˚, 50˚, -63˚, 79˚) have been measured near the frequencies where the topological transition occurs, and experimentally measured field distributions and dispersion bands are shown in Fig. 3. Topological transitions at the magic angles with flattened dispersion can be clearly observed, offering a new avenue to tune photonic topological transitions and topologically engineer the polariton dispersion, different from previous approaches based on sweeping the excitation frequency, which is typically associated with resonant material damping at the transition frequency[31].

Near the topological transition magic angle, the isofrequency dispersion contour flattens and PhPs are highly collimated, directive and diffractionless, as expected in the canalization regime[30] with nearly fixed group velocity directions ($\vec{v}_g = \nabla_{\vec{k}}\omega$). Canalization supported by such flat bandstructure has previously been explored for non-diffractive wave propagation, which opens exciting opportunities for nanoimaging, radiative energy transfer and enhanced local density of states[30]. However, its practical realization based on metamaterials[30, 33] and photonic crystals[34] has been hindered by loss, bandwidth restrictions and lack of tunability. The interplay of rotation angle and tunable interlayer coupling in our proposed bilayers ideally addresses these issues and provides robust polariton canalization over a wide range of frequencies, which may be aligned with a region of low-loss propagation. As the further experimental evidence, we fabricated multiple edges of the sample for Δθ=-63˚ (see scanning electron microscope image in inset of Fig. 4a), so that we can compare polariton propagation in isolated layers (edge 1 and 3, Fig. 4a) and in bilayers (edge 2, 4, 6, 7). As shown in Fig. 4a, the measured PD-scattered PhPs in the bilayer are highly collimated along the same group velocity direction (red arrows). The tip-launched and edge-reflected PhPs form standing-wave patterns near the edge, all showing the same group velocity independent of the edge orientation. As a comparison, edge 1 (E1) on the bottom α-MoO$_3$ and edge 2 (E2) on the tBL α-MoO$_3$ flakes lie parallel along the *x*-axis. However, E1 shows a hyperbolic expanding wavefront (the zoom-in pattern, Fig. 4b), while E2 shows a diffractionless collimated beam (the zoom-in pattern, Fig. 4c). Similar phenomena can be observed comparing edge 3 (E3, Fig. 4d) and edge 4 (E4, Fig. 4e). Even more strikingly, the PhP fringes are not observed parallel to E3 in the single layer, but they are indeed clearly parallel to E4 in tBL α-MoO$_3$, as a result of the extreme dispersion modification enabled by the twist. By carefully examining the PhPs at the end of edges 5, 6, and 7 in Fig. 4a, we find that PhPs consistently propagate along the red arrows, i.e., the group velocity reference direction. These features confirm and corroborate our analytical and numerical results predicting a flat band at ω=903.8 cm$^{-1}$, which corresponds to a fixed group velocity direction independent of the phase velocity (Fig. 4f).

To further demonstrate the diffractionless nature of PhPs at the transition angle, we take a line plot of the measured signal near the PD perpendicular to the group velocity direction, as indicated by the purple lines in Fig. 4a. The results (red dots, Fig. 4g) and the fit by a Gaussian envelope (purple solid line, Fig. 4g) provide a full-width-at-half-maximum (FWHM) around 280 nm (~$\lambda_0$/40, $\lambda_0$ is the free-space wavelength), showing a deeply subwavelength PhP confinement. Such super-resolved canalization PhP mode can be exploited for hyperlensing[30, 33, 34]. In order to study the propagation loss at the observed flattened band, the trajectory plot of the measured PhP signals along the blue line (parallel to the group velocity direction) is shown in Fig. 4f. After fitting, the measured exponential decay length is ~2.77 μm with an inverse damping ratio ~17, while we numerically calculated that PhP canalization at the isolated layer resonance has an exponential decay length ~0.35 μm (see Fig. S7 and supplementary information) [35]. Thus, our experiments confirm that PhPs in tBL vdW materials offer unique opportunities to sustain topologically robust, diffractionless and low-loss canalization at the topological transition magic angles, stemming from the rotation-controlled interlayer coupling.

To conclude, in this work we have theoretically and experimentally demonstrated topologically robust and highly controllable PhPs in tBL α-MoO$_3$ flakes. By controlling the twisting angle in natural vdW materials, we achieved extreme dispersion engineering, offering a powerful platform to investigate twistronics for polaritons. We have observed polariton topological transitions, going from hyperbolic to elliptic dispersion over a broad frequency range, and directly linked the topological nature of their dispersion with $N_{ACP}$ in reciprocal space, as a result of strong hybridization of the dispersion of each layer, controlled by the rotation angle between two α-MoO$_3$ flakes. A flat polariton band necessarily emerges at a topological transition, supporting highly collimated diffractionless and low-loss polariton propagation and a tunable PhP canalization mode. Our studies provide an interesting avenue to translate recent advances in moiré physics and topological engineering in low-dimensional electronic materials to photonics and polaritonics. Our findings also reveal a route for twist-controlled nanoscale light-matter interactions in natural vdW materials and metamaterials[36, 37], with exciting potential for imaging, sensing, radiative energy control and quantum nanophotonics[38].

**Methods [Redacted]**

**Theoretical Modelling**

**Numerical simulations**
**Sample preparation**
**s-NSOM measurements**

**Supplementary Information** is available in the online version of the paper.

**Acknowledgments**

We thank Dr. Z. Q. Xu for constructive discussion and effort on the sample fabrication. A. A. and G. H. acknowledge support from the Air Force Office of Scientific Research with MURI grant No. FA9550-18-1-0379, the Office of Naval Research, and the National Science Foundation. Q. O. acknowledges the support from ARC Centre of Excellence in Future Low-Energy Electronics Technologies (FLEET). J. W. acknowledges the A*STAR Pharos Funding from the Science and Engineering Research Council (Grant No. 152 70 00015). Z. D. acknowledges the National Natural Science Foundation of China (grant No. 51601131). Q. B. acknowledges the support from Australian Research Council (ARC, FT150100450, IH150100006 and CE170100039). C.-W. Q. acknowledges financial support from A*STAR Pharos Program (grant number 15270 00014, with project number R-263-000-B91-305).

**Author Contributions**

G. H. and Q. O. contributed equally to this work. G. H., Q. B., C. W. Q. and A. A. conceived the idea. G. H., C. W. Q. and A. A. developed the theory, with inputs from Y. M. and A. K.. G. H. performed the simulations. Q. O. and Q. B. led the experiments. G. H. and Q. O. designed the experiments. Q. O. prepared samples and performed the measurement. G. S. and Q. O. fabricated the defects and edges on the sample. Y. W. contributed to material synthesis. G. H., Q. O., J. W., Z. D., Q. Z., Q. B., C. W. Q. and A. A. analyzed the data and all authors discussed the results. A. A., Q. B., C. W. Q. supervised the projects. G. H. initiated the draft with inputs and comments from all authors.

**Competing interests**

The authors declare no competing interests.

**Data availability**

The data that support the findings of this study are available from the corresponding author upon reasonable request.

**Code availability**

The codes that support the findings of this study are available from the corresponding author upon reasonable request.

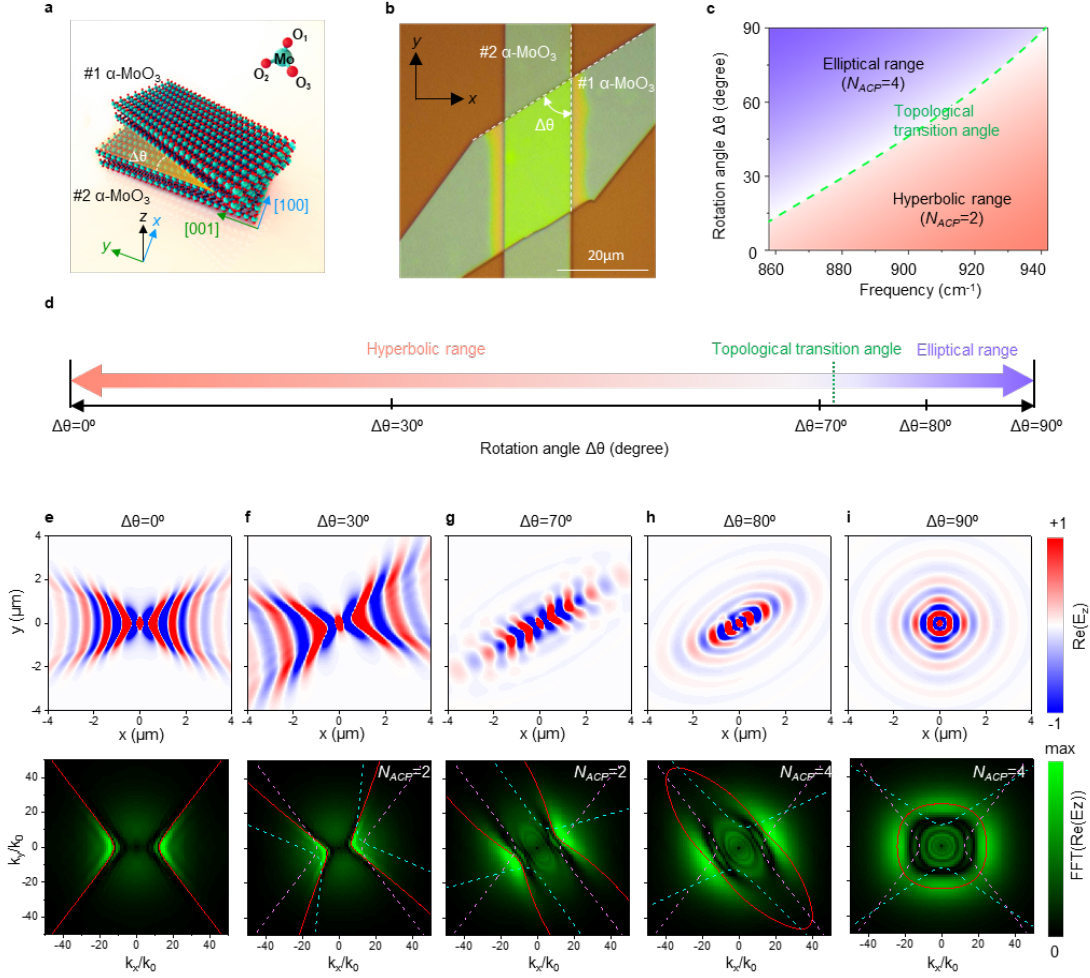

**Fig. 1 Rotation-induced topological transition of phonon polaritons. a**, Schematic of the twisted bilayer (tBL) α-MoO$_3$. Top layer (#1 α-MoO$_3$) and bottom layer (#2 α-MoO$_3$) have thickness $d_1$ and $d_2$, respectively. The $x$ and $y$ axes are along [100] and [001] directions of #2 α-MoO$_3$. The twisted angle is defined as anti-clockwise rotation of #1 α-MoO$_3$ with respect #2 α-MoO$_3$. **b**, Optical image of a tBL α-MoO$_3$ flake (Δθ=-57˚, $d_1$=$d_2$=150 nm). The white dashed lines denote the [001] crystal directions of the two layers. **c**, Map of the topological nature of PhP dispersion with respect to the rotation angle and frequency. The dashed green line corresponds to the topological transition angle (θ$_{TP}$), which separates hyperbolic (red gradient, Δθ <θ$_{TP}$ for $N_{ACP}$=2) and elliptic dispersion regimes (blue gradient, Δθ >θ$_{TP}$ for $N_{ACP}$=4). **d**, Topological nature of the polariton dispersion with respect to the rotation angle at the frequency of 925.9 cm$^{-1}$. **e-i**, Numerically simulated field distributions [top panel, Re($E_z$)] and corresponding dispersion [bottom panel, Fourier transform of (Re($E_z$))] at 925.9 cm$^{-1}$. The red solid lines show the analytically calculated dispersion bands. The magenta dashed and cyan dashed lines correspond to the dispersion curves of the isolated bottom (no rotation) and top (with rotation) α-MoO$_3$ flakes. The bilayer dispersions for Δθ=30˚ and 70˚ are hyperbolic as Δθ<72˚ and $N_{ACP}$=2 and for Δθ=80˚ and 90˚ are elliptic as Δθ>72˚ and $N_{ACP}$=4.

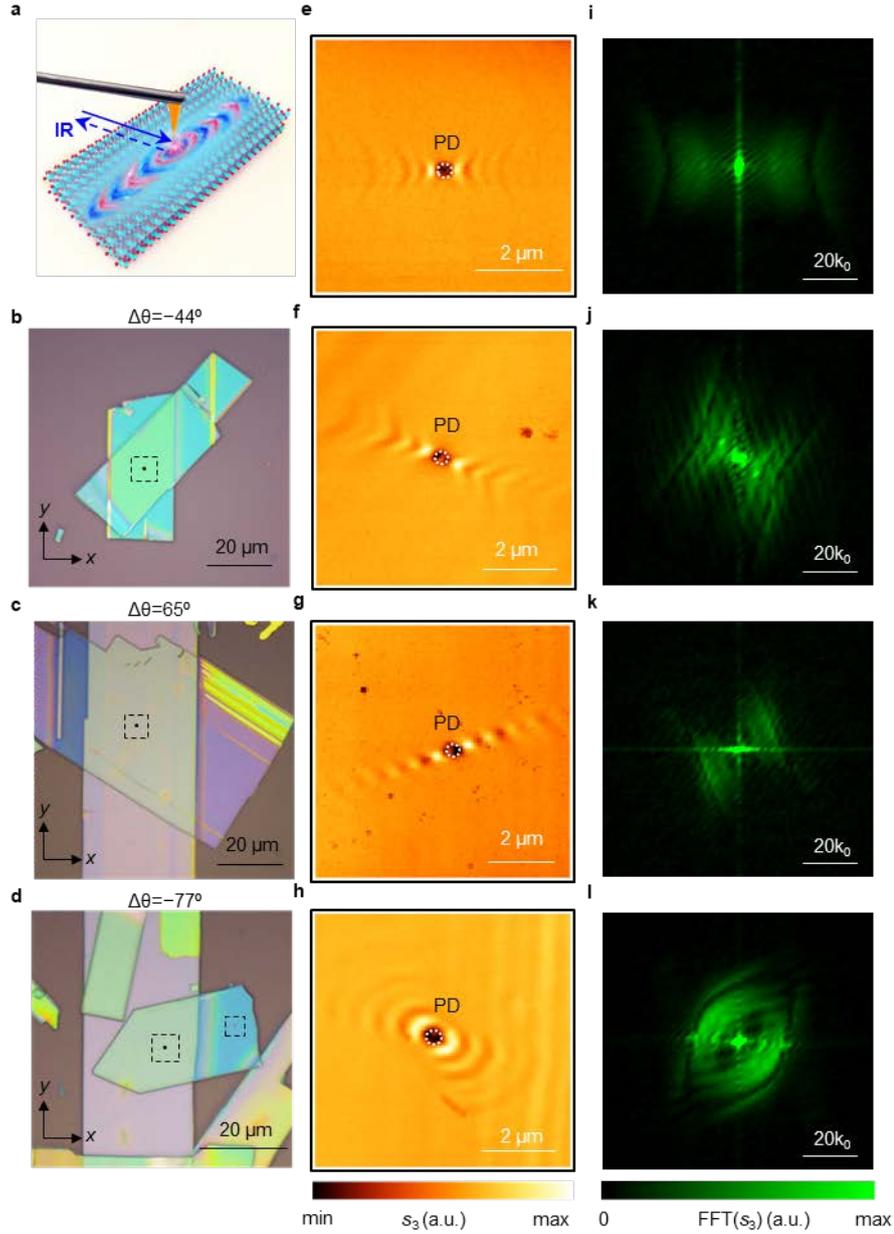

**Fig. 2 Experimental observation of topological polaritons. a**, Schematics of the s-SNOM nanoimaging of PhPs. Upon infrared (IR) light illumination, the metallic tip launches the polaritons in the sample, which are then scattered into free space for collection. **b-d**, Optical images of tBL α-MoO$_3$ samples for Δθ=-44° ($d_1$=$d_2$=128 nm), Δθ=65° ($d_1$=120 nm, $d_2$=235 nm) and Δθ=-77° ($d_1$=125 nm, $d_2$=210 nm). **e-h**, Experimentally measured near-field distribution ($s_3$) near a point defect (PD) as denoted by white dashed circles: **e** is measured near the PD in the top layer of sample with Δθ=-77° (the top dashed square areas in panel **d**); **f**, **g**, and **h** are measured in the tBL α-MoO$_3$ flake region in the dashed square areas of panel **b**, **c**, and **d**, correspondingly. The diameter of PD is around 200 nm. **i-l**, Dispersion curves obtained by Fourier transforming the measured near-field signal ($s_3$) in **e**, **f**, **g** and **h**, respectively. All measurements refer to ω=903.8 cm$^{-1}$. The black dots in panel **b**, **c**, and **d** represent the PD.

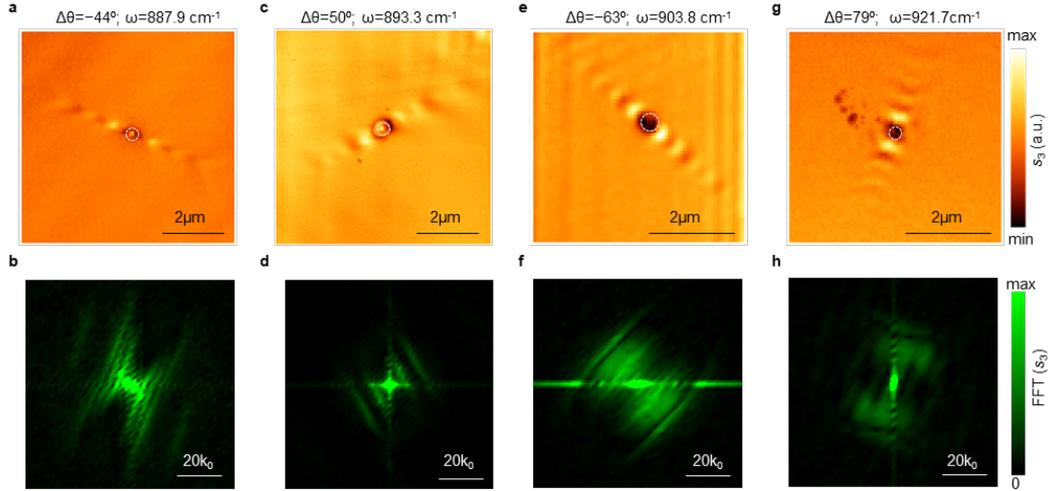

**Fig. 3 Tunable topological transition magic angles. a, b,** Near-field image and corresponding flattened dispersion via Fourier transform at ω=887.9 cm$^{-1}$ for tBL α-MoO$_3$ flakes (Δθ=-44˚, $d_1$=$d_2$=128 nm). **c, d,** Near-field image and flattened dispersion at ω=893.3 cm$^{-1}$ in the sample with Δθ=50˚, $d_1$=120 nm, $d_2$=117 nm. **e, f,** Near-field image and flattened dispersion at ω=903.8 cm$^{-1}$ in the sample with Δθ=-63˚, $d_1$=$d_2$=117 nm. **g, h,** Near-field image and flattened dispersion at ω=921.7 cm$^{-1}$ in the sample with Δθ=79˚, $d_1$=223 nm, $d_2$=213 nm. The point defects are illustrated as the white dashed circles in the near-field image and have the diameter around 200 nm. All near-field images are measured near the topological transition magic angles, showing diffractionless polariton propagation and flattened dispersion.

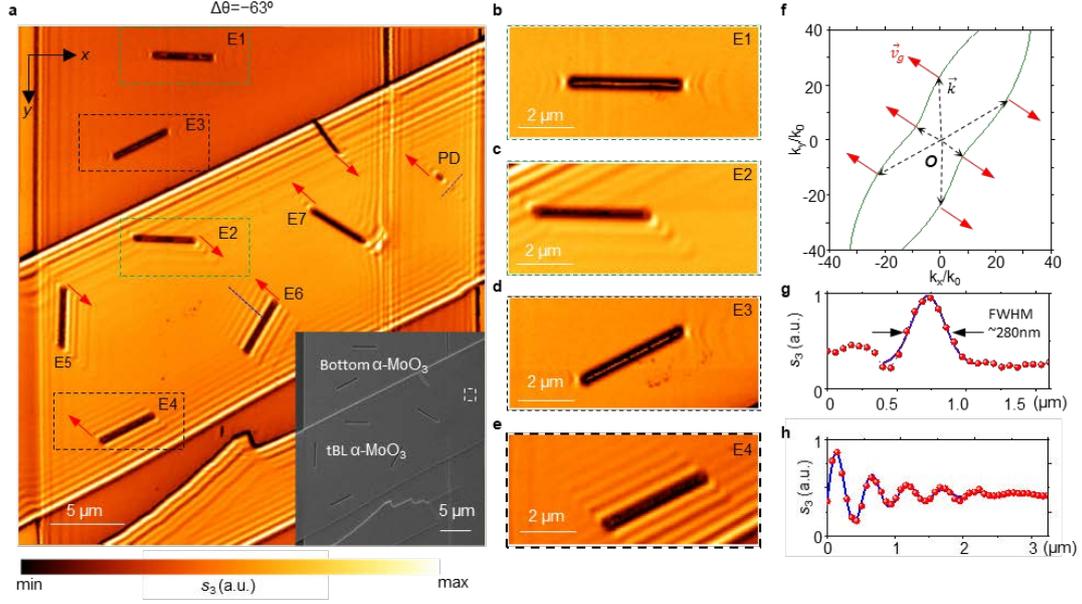

**Fig. 4 Polariton canalization near the topological transition magic angle. a**, Near-field images ($s_3$) of the sample. Seven edges (abbreviated as "E") and one PD are fabricated: E1 and E2 are parallel along the [100] crystal axis of the bottom layer; E3 and E4 are parallel along the [001] crystal axis of the top layer. E5 is along the [001] crystal axis of the bottom layer. The red arrows correspond to the calculated group velocity. The inset shows a scanning electron microscope image of the sample (Δθ=-63°, $d_1$=$d_2$=117 nm). Edges with different orientation and PD (inside the white square) are fabricated in a high-quality large-area sample. **b-e**, Zoom-in measured signals near E1, E2, E3 and E4, respectively. **f**, Analytical dispersion of the sample at ω=903.8 cm$^{-1}$. Black dashed lines represent the polariton wavevectors and red solid lines denote the group velocity directions, which are parallel to each other. **g**, Line plot of measured PhPs along the purple line in panel **a** near the PD. Red dots are the measured signal and the purple solid line is the Gaussian fitting. **h**, Line plot of the signal along the blue trajectory in panel **a** near E6. Red dots are the measured signals and the blue solid line is the fitting.